\documentstyle[proc]{rspublic}

\begin{document}

\def\lap{\lower.5ex\hbox{$\; \buildrel < \over \sim \;$}}
\def\gap{\lower.5ex\hbox{$\; \buildrel > \over \sim \;$}}

\input psfig

\title[Issues for Surveys]{Issues for the Next Generation of
Galaxy Surveys} 

\author[P. J. E. Peebles]{P. J. E. Peebles}

\affiliation{Joseph Henry Laboratories, Princeton University,\\
Princeton NJ 08544, United States of America\\
pjep@pupgg.princeton.edu}

\maketitle

\label{firstpage}

\begin{abstract}
I argue that the weight of the available evidence favours the
conclusions that galaxies are unbiased tracers of mass, the mean
mass density (excluding a cosmological constant or its
equivalent) is less than the critical Einstein-de~Sitter value,
and an isocurvature model for structure formation offers a viable
and arguably attractive model for the early assembly of galaxies. If
valid these conclusions complicate our work of adding structure
formation to the standard model for cosmology, but it seems
sensible to pay attention to evidence.
\end{abstract}

\section{Standard Models and Paradigms}

In cosmology we attempt to draw large conclusions from limited
and often ambiguous data. I am impressed at how well the
enterprise is succeeding, to the point that we have an
established standard model for the hot expanding universe
(Peebles {\it et al.} 1991). Which elements to include in the
standard model is a matter for ongoing debate, of course. I am
inclined to 
take a conservative line if only to avoid giving misleading
impressions to our colleagues with deconstructionist tendencies.
For example, the adiabatic cold dark matter model for structure
formation has been more successful than I expected, and as a
result is rightly the model most commonly used in studies of how 
structure might have formed. Simon White calls this model a
paradigm, which I take to mean a pattern many find useful and
convenient in their research. I like this use of the term,
provided we agree to distinguish it from a well-established
standard model.\footnote{In Kuhn's (1970)  
picture ``normal science'' is done within the framework of a
paradigm. I hope we can agree that those of us who study models
for structure formation outside the set of ideas in the
adiabatic cold dark matter paradigm are not necessarily doing
abnormal science.} I think we cannot count the
adiabatic cold mark matter paradigm as part of the standard model
for cosmology because, as argued here, there is a viable and
perhaps even more attractive alternative.

I organize this discussion around the issues of the
weight of the evidence on whether galaxies are good tracers of
mass, what we are learning from the cosmological tests, and
the elements of the standard model for structure formation
on the scale of galaxies and larger. I begin with another
question, whether Einstein's introduction of the
cosmological principle set a good example for research in our
field. The hurried reader will find the main points summarized in
\S~6.

\section{The Cosmological and Biasing Principles}

The tension between caution and adventure in the
advance of science is well illustrated by the
histories these two principles.

Einstein (1917) introduced modern cosmology with his application of
general relativity theory to a universe that is spatially
homogeneous on average (that is, a stationary random process).
Milne gave the homogeneity assumption its name, Einstein's
cosmological principle. It
is difficult to find in the published literature evidence that
Einstein was aware of the observational situation on the
distribution of matter. Astronomers
had established that we live in a bounded island universe of
stars, and some had speculated that the spiral nebulae are other
island universes. De~Sitter (1917) was willing to consider the
possibility that the nebulae are uniformly distributed in the
large-scale mean, and that their mass constitutes Einstein's
near-homogeneous world matter.\footnote{Published comments
suggest de~Sitter 
considered Einstein's ideas on this point somewhat speculative,
while Einstein felt that de~Sitter's conservative attitude was a
little defeatist. It will be fascinating to see whether the
de~Sitter archives yield any letters exchanging views on the 
cosmological principle.} 
On the other hand, de~Sitter was well aware that the distribution
of the nearby nebulae is decidedly clumpy; indeed, Charlier (1922)
pointed out that it resembles a clustering
hierarchy (what we would now call a fractal). That is, the
conservative advice from the astronomical community would have
been that the observations do not support Einstein's world
picture, that he would do well to consider a fractal model
instead. 
But now Einstein's cosmological principle is well established and
part of the standard model: fluctuations from homogeneity on the
scale of the Hubble length are less than one part in $10^3$ (from
the isotropy of the X-ray background, and about one part in
$10^4$ in the standard relativistic model; Peebles 1993). This
is a magnificent triumph of pure thought!

Just as the cosmological principle was introduced by hand to
solve a theoretical problem, the violation of Mach's
principle in asymptotically flat spacetime, the biasing principle
was introduced to reconcile the low relative peculiar velocities
of the galaxies with the high mass density of the theoretically
preferred Einstein-de~Sitter world model 
(Davis {\it et al.} 1985). There never has been any
serious observational evidence for biasing, 
but the idea rightly was taken seriously because it
is elegant and plausible.\footnote{The 
different distributions of early-type galaxies, of spirals,
and of starburst galaxies cannot all trace the mass, and it has
been very correctly noted that in this sense biasing manifestly
obtains (e.g. Guzzo {\it et al.} 1997). But the spheroid
components of the galaxies seem to be 
the most robust against environment-dependent effects such as
mass loss --- biasing through evolution rather than birth --- and
my  impression is that within the uncertainties all 
dynamical analyses are consistent with the assumption that the
spheroid light traces the mass on the scale of inter-galaxy
distances. And optical samples are reasonable tracers of the
spheroid component.} But I do not include biasing 
in the standard model; we have no very strong evidence for it and
the following three arguments against it.

First, there is no identification of a population of void irregular
galaxies, remnants of the assumed suppression of galaxy formation
in the voids (Peebles 1989). The first systematic redshift survey
showed that the distributions of low and high luminosity galaxies
are strikingly similar (Davis {\it et al.} 1982). I know of no
survey since, in 21-cm, infrared, ultraviolet, or low surface
brightness optical, that reveals a void population. There is a 
straightforward interpretation: the voids are nearly
empty because they contain little mass.

Second, the improving suite of cosmological tests
listed in the next section suggests the mean mass density
is well below the Einstein-de~Sitter value. If the density is low
it means galaxies move slowly because there is not much mass
to gravitationally pull them, not because they are biased tracers
of the mass. 

Third, the galaxy autocorrelation function at low redshift has a
simple form, quite close to the power law
$\xi _{\rm gg}\propto r^{-\gamma}$, with $\gamma =1.77\pm 0.04$,
over three orders of magnitude in separation, 
$10\hbox{ kpc}\lap hr\lap 10\hbox{ Mpc}$. Carlberg shows in these
Proceedings that the index $\gamma$ is quite close to constant back to
redshifts near unity. On the theoretical side, Simon White
describes elegant numerical 
simulations of the adiabatic CDM model. In these simulations the
mass autocorrelation function $\xi _{\rho\rho}(r)$ is not close
to a power law, and the slope of $\xi _{\rho\rho}(r)$ increases
with increasing time. The two functions allow us to define a bias
parameter,
\be
  b(r,t) = \left[ \xi _{gg}(r,t)/\xi _{\rho\rho}(r,t)\right] ^{1/2}.
\label{eq:bias}
\ee
In the adiabatic CDM model this is a function of separation and
time. One interpretation is galaxies are
biased tracers of mass, the bias depending on scale
and time. But why should the biased tracer exhibit a striking
regularity, in $\xi _{gg}(r)$ and the three-and four-point
functions, that is not a property of the mass that is driving
evolution? The more straightforward reading is that the regularity
in $\xi _{gg}(r)$ reflects a like regularity in the behaviour of
the mass, and that there is a slight flaw in the model. Given the 
enormous step we are taking in analyzing the growth of the
structure of the universe it surely would not be surprising to
learn that we have not yet got it exactly right. 

\section{The Cosmological Tests}

\subsection{The Purpose of the Tests}

In the standard Friedmann-Lema\^{\i}tre cosmological model
coordinates can be assigned so the mean line element is
\be
	ds^2 = dt^2 - a(t)^2\left[ {dr^2\over 1\pm r^2/R^2}
	+ r^2(d\theta ^2 + \sin ^2\theta d\phi ^2)\right] .
\ee
The mean expansion rate satisfies the equation
\be 
	H^2 = \left( \dot a\over a\right) ^2 
	= {8\over 3}\pi G\rho \pm {1\over a^2R^2} + 
	{\Lambda\over 3},
\ee
which can be approximated as
\be
	H^2= H_o^2[\Omega (1+z)^3 + \kappa (1+z)^2 +\lambda ].
\label{eq:cos_pars}
\ee
This defines the fractional contributions to the
square of the expansion rate by matter, space curvature, and the
cosmological constant (or a term in the stress-energy tensor that
acts like one). The time-dependence assumes pressureless matter
and constant $\Lambda$. 
Other notations are in the literature; one that
is becoming popular adds the matter and $\Lambda$ terms, as in
Michael Turner's contribution to these Proceedings. To avoid
confusion with the definitions in equation~(\ref{eq:cos_pars}) we
might express Turner's convention as
\be
	\Omega ' = \Omega + \lambda.
\ee
This isolates the curvature term, which is useful. 
And since the evidence is that $\Omega$ is small it certainly
helps rescue our theoretical preference for a density parameter
equal to unity. I find it unsatisfying, 
however: what became of the intense debates we had on biasing
and the other systematic errors in the measurement of $\Omega$? 

I can get more excited about the Full Monty: let
\be
	\Omega ^{\prime\prime } = \Omega +\kappa +\lambda 
	= 1 - \kappa .
\label{eq:fm}
\ee
Each of the terms on the right-hand side of this equation is
measurable in principle, and if the applications of the
cosmological tests continue to improve at the present rate it may
not be many more years before we have ten percent measurements of
the three numbers. If they add to unity we will have a test of
general relativity theory applied on large scales in the strong
curvature limit.

%\begin{figure}
%\centerline{\psfig{file=table1.ps,width=3.75truein,clip=}}
%\end{figure}

{\it Here's my problem: the conference wants LATEX, the table is
set in plain TEX because it's much too fiddly for LATEX, and LANL
won't accept a postscript file of the compiled file of the
table. What is a computer-challenged person to do?}

The point is illustrated another way in
table~1. The lines represent quite different ways to probe the
standard relativistic model, and the columns are grades for
how well three sets of parameter choices fit the results. As the 
observations improve we may find that only one
narrow range of parameters is consistent with all the
constraints. If so we will have settled two issues. 

First, it surely will continue to be difficult to use internal
evidence to rule out systematic errors in astronomical
observations. For example, can astronomers unambiguously
demonstrate that SNeIa in a given class of light curve shape
really are drawn from the identical 
population at redshifts $z \sim 0$ and $z\sim 1$? A consistent
story from independent tests is strong evidence the measurements
have not been corrupted by some subtle systematic error. 

Second, a consistent story will be a strong positive test of the
standard relativistic cosmological model, as in
equation~(\ref{eq:fm}).
The successful parameter set could be quite different from any of  
the choices in the table, of course; we may be driven to a
dynamical $\Lambda$, for example (Peebles \&\ Ratra 1988; 
Huey {\it et al.} 1998).

The classical cosmological tests based on measures of the
spacetime geometry have been supplemented by a new class of
tests based on the condition that the cosmology admit a
consistent and observationally acceptable model for structure
formation (categories 3 and 4 in table~1). I comment on
some aspects of structure formation in \S 4 and \S 5.

\subsection{The State of the Tests}

The constraint from the rate of lensing of quasars by foreground
galaxies does not comfortably fit the 
curvature of the redshift-angular distance relation. The analysis
of lensing by Falco {\it et al.} (1998), for a combined sample of
lensing events detected in the optical and radio, indicates that
if the universe is cosmologically flat then $0.64<\Omega <1.66$
at one standard deviation, and $\Omega > 0.38$ at $2\sigma$. The
SNeIa redshift magnitude relation, from the magnificent work of 
Perlmutter {\it et al.} (1998) and Reiss {\it et al.} (1998),
seems best fit by $\Omega = 0.2$, $\lambda = 0.8$. The
discrepancy is not far outside the error flags, but I think that
if the lensing rate were the only available cosmological test we
would greet it as confirmation of the Einstein-de~Sitter model
and another success for pure thought. 

The lensing constraint depends on the galaxy mass function.
The predicted peak of the lensing rate at angular separation
$\theta\sim 1$ arc~sec is dominated by the high surface density  
branch of early-type galaxies at luminosities $L\sim L_\ast$. 
The number density of these objects is not well known, and an
improved measurement is an important goal for the new generations
of surveys of galaxies. If further tests of the lensing and
redshift-magnitude constraints confirm an inconsistency for
constant $\Lambda$ the 
lesson may be that the cosmological constant is dynamical,
rolling to zero, as Ratra \&\ Quillen (1992) point out. 

The Einstein-de~Sitter model is not yet ruled out, but I 
think most of us would agree that consideration of
structure formation in low density cosmological models is well
motivated.

\section{The Origin of Large-Scale Structure}

We have good reason to think galaxies grew by gravity out of
small initial departures from homogeneity, but the nature of the
initial conditions is open to discussion. To illustrate this I
present some elements of an isocurvature model. Details are in
Peebles (1998$a$, $b$). 

\subsection{Adiabatic and Isocurvature Models}

In the paradigm Simon White describes in these Proceedings
structure grows out of an adiabatic departure from homogeneity 
--- as would be produced by local reversible expansion or
contraction from exact homogeneity --- that is a spatially
stationary isotropic random Gaussian process. Another possibility
is that the primeval mass distribution is exactly homogeneous ---
there is no perturbation to spacetime curvature --- and structure
formation is seeded by an inhomogeneous composition. In the
isocurvature model presented here the initial entropy per
baryon is homogeneous, to preserve the paradigm for element
formation, and homogeneity is broken by the distribution
of cold dark matter. In both models the present mass of the 
universe is dominated by nonbaryonic cold dark matter (CDM); I 
shall call them ACDM and ICDM models. 

\subsection{Power Spectra}

In the ACDM model the primeval mass density fluctuation
(defined as the most rapidly growing density
perturbation mode in time-orthogonal coordinates) has a close to
power law power spectrum, $P\propto k^n$. 
In the ICDM model the primeval distribution
of the CDM is close to a power law, $P\propto k^m$, in a
homogeneous net mass distribution. It is an interesting exercise
to check that in linear perturbation theory the evolution from
the initial radiation-dominated universe to the present
CDM-dominated epoch bends the spectra to 
\begin{eqnarray}
&& \hbox{ACDM:}\qquad  P\propto k^{n - 4},\ k\gg k_{\rm eq},\qquad
	P\propto k^n,\ k\ll k_{\rm eq},\label{eq:Aspect}\\
&& \hbox{ICDM:}\qquad  P\propto k^m,\ k\gg k_{\rm eq},\qquad
	P\propto k^{m + 4},\ k\ll k_{\rm eq},\label{eq:Ispect}
\end{eqnarray}
where $k_{\rm eq}$ is the wavenumber appearing at the Hubble
length at the redshift $z_{\rm eq}$ of equality of mass densities
in matter and radiation. 

\begin{figure}
\centerline{\psfig{file=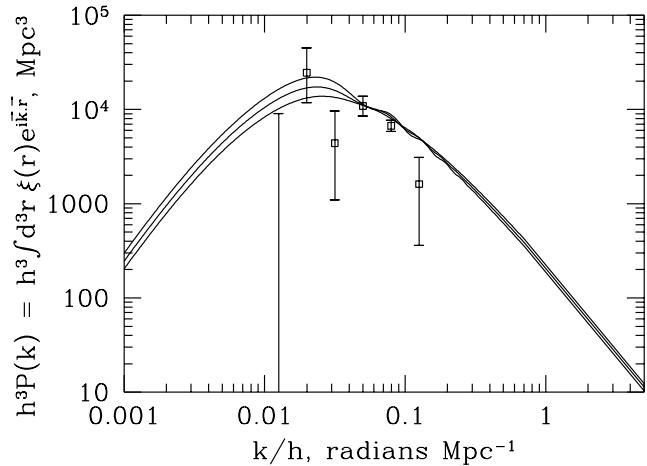,width=3.5truein,clip=}}
\caption{Power spectrum of the CDM space distribution in the ICDM model
at the present epoch computed in linear perturbation theory for
the parameters in equations~(\ref{eq:parameters})
and~(\ref{eq:Pnorm}). The density parameter in 
baryons is $\Omega _B=0.05$ in the top curve at small $k$, 0.03
in the middle, and 0.01 in the bottom curve. The data are from
the PSC-z survey (Saunders {\it et al.} 1998).} 
\end{figure}

The similarity of equations~(\ref{eq:Aspect}) and~(\ref{eq:Ispect}) 
for $m \sim n - 4$ extends to roughly similar
spectra of the angular distribution of the thermal cosmic
background radiation (the CBR) in the adiabatic and isocurvature
CDM models. The status of ACDM model fits to the fluctuation
spectra of galaxies and the CBR is discussed in
these Proceedings by Bond. Figures~1 and ~2 show the 
ICDM model predictions for the parameters
\be
	m = -1.8,\qquad\Omega = 0.2, \qquad\lambda = 0.8, 
	\qquad h = 0.7, \label{eq:parameters}
\ee
with the normalization
\be
	P(k) = 6300h^{-3}\hbox{ Mpc}^3 \hbox{ at } 
	k = 0.1h\hbox{ Mpc}^{-1},
\label{eq:Pnorm}
\ee
where Hubble's constant is $H_o=100h$~km~s$^{-1}$~Mpc$^{-1}$. 
The data in figure~1 are from the IRAS PSC-z (point source
catalog) redshift survey of Saunders {\it et al.} (1998). This is
the real space 
spectrum after correction for peculiar velocity distortion
represented by the density-bias parameter $\beta =0.6$. There are
good measurements of the 
spectrum of the galaxy distribution on smaller scales,
$k>0.1h$~Mpc$^{-1}$, but this approaches the nonlinear sector,
and it seems appropriate to postpone discussion of the
small-scale mass distribution until we have 
analyses of nonlinear evolution from the non-Gaussian initial
conditions of the model in equation~(\ref{eq:ICDMrho}). Since the
PSC-z catalog is deep, with 
good sky  coverage, it promises to be an excellent probe of the
large-scale galaxy distribution, and it is a very useful
normalization. 

\begin{figure}
\centerline{\psfig{file=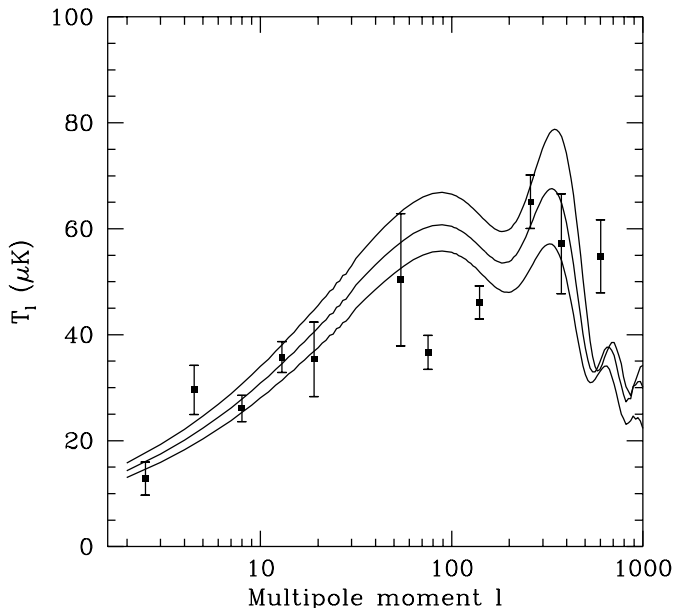,width=3.5truein,clip=}}
\caption{Angular fluctuation spectrum of the CBR in the ICDM model
with the parameters in equations~(\ref{eq:parameters})
and~(\ref{eq:Pnorm}). The density parameter in 
baryons is $\Omega _B=0.05$ in the top curve, 0.03 in the middle,
and 0.01 in the bottom curve. The ionization history is computed
under the assumption that there is no source of ionizing
radiation apart from the CBR.} 
\end{figure}

Figure~2 shows second moments of the angular distribution of the
CBR, where 
\be
	T(\theta ,\phi) = \sum a_l^m Y_l^m(\theta ,\phi ),\qquad
	T_l = \left[ {l(2l + 1)\over 4\pi } \right] ^{1/2}
	\langle |a_l^m|^2\rangle ^{1/2}.
\label{eq:Tl}
\ee
In the approximation of the sum over $l$ as an integral 
the variance of the CBR temperature per 
logarithmic interval of $l$ is $(T_l)^2$.\footnote{There are good 
historical reasons, dating from the introduction of the ACDM
model, for writing $2l(l+1)$ in place of $l(2l+1)$, as does Bond
in his contribution to these Proceedings, 
but since I am considering ICDM the 
convention in equation~(\ref{eq:Tl}), which I prefer 
because it reflects the $2l+1$
components for each value of $l$, may not be unreasonable.}
The measured $T_l$ are from the
compilation of Ratra (1998). 

The second moments of the large-scale distributions of mass and
radiation in the ICDM model agree with the data as well about as
could be expected given the state of these difficult
measurements. The same is true of ACDM models 
considered by Bond; both cases pass. At most one
will pass the improved measurements expected from work in
progress, but that is for the future. 

\subsection{Inflation-Based Model for Isocurvature Initial
Conditions}

Simple and arguably natural realizations of the inflation concept
lead to adiabatic initial conditions; others to isocurvature
initial conditions.  In the example of the latter in Peebles
(1998$a$) the CDM is a scalar field $\phi$ that ends up after
inflation in a squeezed state as a Gaussian random process with
mass density      
\be
	\rho ({\bf x}) = M^2\phi ({\bf x})^2/2,
	\label{eq:ICDMrho}
\ee
for field mass $M$. In a simple case the field satisfies
\be
	\langle\phi\rangle = 0,\qquad
	\langle\phi ({\bf x}_1)\phi ({\bf x}_2)\rangle\propto
	x_{12}^{-\epsilon}, 
\label{eq:ICDMe}
\ee
and the power spectrum of the mass distribution in equation
(\ref{eq:ICDMrho}) is a power law with index $m=2\epsilon -3$. 
The model requires $m = -1.8$, or $\epsilon = 0.6$. The ``tilt''
from the scale-invariant case $\epsilon\simeq 0$ is not difficult
to arrange; whether it might be considered natural has yet to be
debated. 

The primeval density fluctuations in the model in
equations~(\ref{eq:ICDMrho}) and~(\ref{eq:ICDMe}) are
non-Gaussian and scale-invariant: the frequency distribution of
the density contrast $\delta$ averaged through a window and
scaled by the standard deviation 
$\langle\delta ^2\rangle ^{1/2}$ is independent of
the window size. The evidence discussed in Peebles (1998$b$)
indicates the model with these initial conditions is viable but
subject to serious tests from improvements from observational
work in progress. The same is true of the ACDM models, of course.
I turn now to one of the tests, the redshift of assembly of the
galaxies.  

\section{The Epoch of Galaxy Assembly}

\subsection{Scaling Galaxies from Clusters of Galaxies}

The power law model for the primeval CDM fluctuation
spectrum (eqs.~[\ref{eq:Ispect}] and~[\ref{eq:parameters}])
is a good approximation for the residual CDM mass distribution at
redshifts less 
than the epoch $z_{\rm eq}$ of equality of mass densities in
matter and radiation and on scales small compared to the 
Hubble length at $z_{\rm eq}$ and large
compared to the scale of nonlinear clustering. Within these
bounds the spectrum varies as
\be
	P_\rho\propto k^mD(t)^2,
\label{eq:scaling}
\ee
where $D(t)$ is the solution to the linear equation for the
evolution of the density contrast in an isothermal perturbation
of the CDM. The rms contrast through a window of comoving radius
$x$ varies as  
\be
	\delta\propto x^{-(3+m)/2}D(t).
\ee
Gravitational structure formation
is triggered by passage of upward fluctuations of $\delta$
through unity, and the threshold is not sensitive to $\Omega$ in
a cosmologically flat model. This means the characteristic
physical length, mass, and internal velocity of newly forming
structures scale with time as  
\be
	r_{\rm nl}\propto (1+z)^{-1}D^{2/(3+m)},\quad
	M\propto D^{6/(3+m)}, \quad
	\sigma\propto (1+z)^{1/2}D^{2/(3+m)}.
\ee
These relations neglect nongravitational interactions; they may
be expected to be useful approximations on scales much larger than the
half-light radii in galaxies, where the CDM halo dominates the
mass in the standard model. 

We can normalize to the great clusters of galaxies, with 
\bea
	r_{\rm A} &=& 1.5h^{-1}\hbox{ Mpc}, \quad
	\sigma _{\rm cl} = 750\hbox{ km s}^{-1}, \nonumber\\
	m_{\rm cl} &=& 4\times 10^{14}h^{-1}M_\odot ,\quad
	n_{\rm cl} =(2\pm 1)\times 10^{-6}h^{-3}\hbox{ Mpc}^{-3}.
	 \label{eq:mcl}
\eea
The Abell radius is $r_{\rm A}$, $\sigma _{\rm cl}$ is an rms mean
line of sight velocity dispersion for $R\geq 1$ clusters, 
$m_{\rm cl}$ is the mean mass within the Abell radius, and 
$n_{\rm cl}$ is the present number density of clusters with mass
$m>m_{\rm A}$ (Bahcall \&\ Cen 1993). Clusters are relaxing
at the Abell radius, and the merging rate is significant, but it
is generally agreed that that internal velocities
typically are close to what is needed for support against
gravity at $r\sim r_{\rm A}$. In the power law model in
equation~(\ref{eq:scaling}) these 
quantities scaled back in time characterize objects in a like
state of early development in the past. 

With the parameters used in figures~1 and~2
(eq.~[\ref{eq:parameters}]) the scaling relations applied at
expansion factor $1+z=7$  give
\be
	r_{\rm g}=15h^{-1}\hbox{ kpc},\qquad
	\sigma _{\rm g}= 140\hbox{ km s}^{-1},
	\qquad M_{\rm g}=1.3\times10^{11}h^{-1}M_\odot.
\label{eq:young_galaxies}
\ee
The present characteristic separation of clusters and the scaled
comoving separation at $1+z=7$ are
\be
	d_{\rm cl}=n_{\rm cl}^{-1/3} = 80h^{-1}\hbox{ Mpc},\qquad
	d_{\rm g} = 5h^{-1}\hbox{ Mpc}.
\ee
In this model an astronomer sent back in time to
$1+z=7$ would see objects with the somewhat disordered appearance
of present-day clusters, merging at a significant rate, but
with internal motions typically close to what is needed for virial
support. The characteristic size, mass, and comoving distance
between objects would be seen to be characteristic of the
luminous parts of present-day $L_\ast$ galaxies. Our time 
traveller might well be inclined to call these objects young
galaxies, already assembled at $z=6$.

At expansion factor $1+z=20$ the scaling relations give
\be
	r\sim 1\hbox{ kpc},\qquad\sigma\sim 40\hbox{ km s}^{-1},
	\qquad m\sim 1\times10^{9}M_\odot ,
\ee
numbers characteristic of dwarf
galaxies. I have to assume many merge to form the $L_\ast$
giants, and that the merging rate eases off at $1+z\sim 7$,
perhaps because the dissipative settling of the baryons has
progressed far enough to lower the cross section for merging, so
later structure formation can build the present-day galaxy
clustering hierarchy. 

If galaxies were assembled as mass concentrations at $1+z=7$, as
this model suggests, how would they appear at $1+z\simeq 4$?
Internal velocities ought to be characteristic of present-day
galaxies. That is not inconsistent with the properties of the
damped Lyman-$\alpha$ absorbers studied by Wolfe \& Prochaska
(1998), though Haehnelt, Steinmentz \&\ Rauch (1998) show other
interpretations are possible. The
expected optical appearance depends on how feedback affects the
rate of conversion of gas to stars, a delicate issue I am
informed. In these Proceedings Steidel presents elegant
optical observations of high redshift galaxies that reveal strong
spatial clustering. Steidel points out this could signify strong
biasing at formation. The interpretation could be slightly
different in the non-Gaussian ICDM model, where high
density fluctuations tend to appear in concentrations (Peebles
1998$c$). 

Structure formation happens later in ACDM. I have expressed
doubts that late assembly could produce the high density
contrasts of normal present-day  $L_\ast$ 
galaxies, but the numerical simulations White describes seem
not to find this a problem. If dense galaxies can be assembled at
low redshift, when the mean mass density is low, one might have 
thought that protogalaxies assembled at high redshift and high 
mean mass density would be unacceptably dense. But Nature 
was able to form clusters of galaxies that are close to virial
equilibrium at modest density contrast at the Abell radius, and
well enough isolated that they seem likely to remain part of the
clustering hierarchy rather than merging into larger monolithic
superclusters. Under the scaling argument the same would be true
of protogalaxies assembled at $1+z\sim 7$ in the ICDM model. 

\subsection{Collapse Models}

I arrived at the isocurvature model in \S 4 (and Peebles
1998$a$ and~$b$) through a search for a model for galaxy
formation at high redshift, when the cosmic mean density is
comparable to that of the luminous parts of a normal large
galaxy. The argument traces back to Partridge \& Peebles 
(1967), a recent version is in Peebles (1998$c$), and elements
are reviewed here.

\begin{figure}
\centerline{\psfig{file=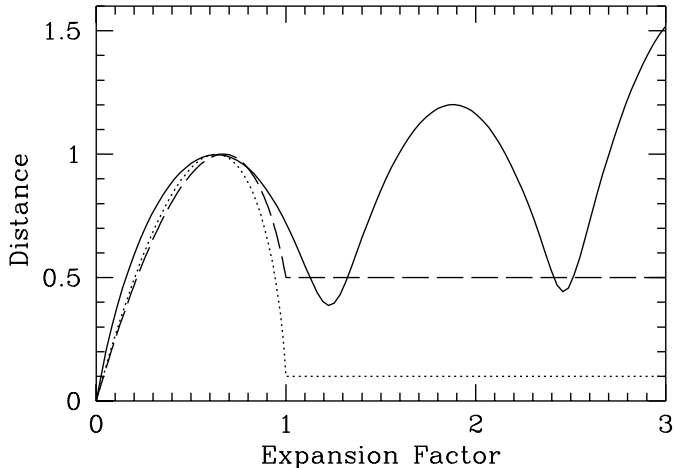,width=3.5truein,clip=}}
\caption{Models for gravitational assembly of galaxies and systems
of galaxies. The solid line is the distance
between the Andromeda Nebula and the Milky Way in a solution for
the interaction with neighboring mass concentrations. For this
curve the present epoch is at expansion parameter $a=1$ and the
distance unit is 0.97~Mpc. The dashed line is a commonly
discussed spherical model. The dotted line models late assembly
of the mass in the central parts of a normal giant galaxy.} 
\end{figure}

The solid line in figure~3 is the distance between the
Andromeda Nebula M31 and our Milky Way galaxy in a numerical solution
for the motions of the galaxies in and near the Local Group
(Peebles 1996). The orbits are constrained to arrive at the
present positions at expansion parameter $a=1$ from initial
motions at $a\rightarrow 0$ consistent with the homogeneous
background cosmological model. This uses the Einstein-de~Sitter
model, so the solution may be scaled with time. 

The dashed line in the figure assumes spherical symmetry with no
orbit crossing, expansion from cosmological initial conditions,
and collapse to half the maximum radius, at which point the
kinetic energy in the spherical model has reached half the
magnitude of the gravitational potential energy. The solid line
for the motion of M31 relative to the Milky Way has a similar
shape but with significant differences. In the numerical
solution neighbouring galaxies are close to the Milky Way and
M31 at $a\lap 0.25$, so the solid line is more strongly curved
than a spherical solution with fixed mass. The solid
line is less strongly curved at larger expansion factor because
the interaction with neighbouring mass concentrations has given
the Milky Way and M31 substantial relative angular momentum. 
The present transverse relative velocity of M31 is 
comparable to the radial velocity of approach, 
the minimum separation is about half the present value, and the
mean separation in the future is larger than the present value.
As we all know, nonradial motions tend to suppress collapse. 

Now let us consider the spherical solution as a model for young
galaxies. Let $r(t)$ be the proper
radius of a sphere that is centred on 
the young galaxy and contains the mass $M_{\rm g}$ in
equation~(\ref{eq:young_galaxies}). In the
spherical solution, which ignores nonradial motion and the motion
of mass across the surface of the sphere, the radius varies with
time as $r=A(1-\cos\eta )$, where $t=B(\eta -\sin\eta )$ and
$A^3=GM_{\rm g}B^2$. This ignores the cosmological constant
$\Lambda$, which has little
effect on the orbit. If spherical collapse stops at radius
$r_{\rm g}$ at redshift $z_{\rm g}$, then in the spherical model
the collapse factor from maximum expansion is  
\be
	r_{\rm g}/r_{\rm max} = (1 - \cos\eta _{\rm g})/2,
\label{eq:cf}
\ee
and an adequate approximation to $\eta _{\rm g}$ is
\be
	{(\eta _{\rm g}-\sin\eta _{\rm g} )^2\over
	(1-\cos\eta _{\rm g})^3} = 
	{8\over 9\Omega }\left(\sigma _{\rm g}
	\over H_or_{\rm g}\right) ^2(1+z_{\rm g})^{-3} =
	{4\times 10^4\over (1+z_{\rm g})^3},
\label{eq:etao}
\ee
for the numbers in equation~(\ref{eq:young_galaxies}). 

For the collapse factor $r_{\rm max}/r_{\rm g}=2$ in the dashed
line in figure~3 equation~(\ref{eq:etao}) says 
$1+z_{\rm g}\sim 10$, not far from the 
value $1+z_{\rm g}=7$ in equation~(\ref{eq:young_galaxies}). 

In a model for late galaxy assembly, at $z_{\rm g}=1$,
equations~(\ref{eq:cf}) and~(\ref{eq:etao}) say 
$r_{\rm rmax}/r_{\rm g}\sim 10$. This is the dotted line in
figure~3. The pronounced collapse could result from 
exchange of energy 
among lumps settling out of a more extended system, as
happens in numerical simulations (Navarro, Frenk, \&\ White
1996), but I think there are two reasons to doubt it happens
in galaxy formation. First, in a hierarchical model for structure
formation collapse to $r_{\rm g}\sim 20$~kpc at $z_{\rm g} =1$
traces back to a cloud of subgalaxy fragments---star
clusters---at radius  $\sim r_{\rm max}\sim 200$~kpc. I know of
no evidence of such clustering (apart from the usual power law
correlation functions) 
in deep samples. Second, the scaled process of formation of rich
clusters of galaxies shows no evidence of pronounced collapse:
clusters seem to be close to stable at the Abell radius and
present in significant numbers at redshift $z=0.5$.

These are arguments, not demonstrations. I consider them
persuasive enough to lend support to the isocurvature model that
leads from the fit to measures of
large-scale structure in figures~1 and~2 to the scaling model for
early galaxy assembly in equation~(\ref{eq:young_galaxies}). 
This depends on whether galaxies really were assembled
early, of course, and are fortunate that the observations Steidel
describes in there Proceedings may well be capable of telling us
when the galaxies formed.

\section{Discussion}

It is inevitable that the exciting rush of advances in this
subject has left ideas unexplored. I have attempted to identify
some roads not taken, less popular lines of thought that seem
worth considering. The main points are summarized in the
following questions.

\subsubsection{Did Einstein set a good example?}

Einstein's brilliant success in establishing key elements of the
standard cosmological model is an example of why we pay serious
attention to elegant ideas even in the face of contrary empirical
indications. But I think this is not an entirely edifying
example: Einstein's intuition was not always so successful,
and most of us are not Einsteins. In the present 
still crude state of cosmology it is better to be led by the
phenomenology from astronomy and from particle physics
(that may teach us the identity of the dark matter, for example). 

Most of us agree that the Einstein-de~Sitter model is the elegant
case, and it makes sense that the community has given it 
special attention despite the long-standing indication from
galaxy peculiar velocities that the Einstein-de~Sitter density is
too high. Now other lines of evidence are pointing in the same
direction, as summarized in table~1, and I think there is 
general agreement in the community that we must give serious
consideration to the possibility that Nature has other ideas
about elegance. I count this as a cautionary example for the
exploration of ideas on how the galaxies formed. 

\subsubsection{Why are galaxies thought to be biased tracers of mass?} 

The galaxy two-point correlation function is quite close to a
power law, $\xi _{gg}(r)\propto r^{-\gamma}$, over three orders
of magnitude of separation $r$ at low redshift, and the index
$\gamma$ is quite close to constant back to redshifts approaching 
unity. This is not true of the mass autocorrelation function 
$\xi _{\rho\rho}(r)$ in the adiabatic cold dark matter (ACDM)
model. Thus we have a measure of bias, 
$b(r,t)=\left[ \xi _{gg}(r,t)/\xi _{\rho\rho}(r,t)\right] ^{1/2}$
(eq.~[\ref{eq:bias}]), that depends on position and time.
Should we take this as evidence galaxies are biased mass
tracers? Since the regularity is in the galaxies surely the first
possibility to consider is that $\xi _{gg}(r)$ is
revealing a like regularity in the behaviour of the mass, that
the bias is in the model. This reading is heavily influenced by a
related issue: if much of the CDM is in the voids defined by 
normal galaxies where are the remnants of the void galaxies?
Surely they are not entirely invisible?  

I am impressed by the elegant simulations of the ACDM models
Simon White presents in these Proceedings, and have to
believe they reflect aspects of reality. But the curious
issue of $b(r,t)$ leads me to suspect there is more to the
story. Would the isocurvature variant do better? That awaits
searching tests by numerical simulations of the kind that that
have been applied to the adiabatic case.  

\subsubsection{What is the purpose of the cosmological tests?}

One often reads that it is to determine how the
world ends. But should we trust an extrapolation into the
indefinitely remote future of a theory that we know can only be a
good approximation to reality? For a trivial example, suppose the universe
has zero space curvature and the present value of the density
parameter in matter capable of clustering is $\Omega = 0.2$, with
the rest of the contribution to $H_o{}^2$ in a term
that acts like a cosmological ``constant'' $\Lambda$ that is
rolling toward zero (Peebles \& Ratra 1988; Huey {\it et al.}
1998). If the final value of $\Lambda$ is identically zero then
the world ends as Minkowksi spacetime (after all the black holes
have evaporated). If $\Lambda$ ends up at a permanent negative
value, no 
matter how close to zero, the world ends in a Big Crunch. Should
we care which it is? I would consider a bare answer an empty
advance, because the excitement of physical science is in
discovering the interconnections among phenomena. Perhaps the 
excitement of knowing how the world ends will be in what it
teaches us about how the world began.

The classical cosmological tests, that probe spacetime
geometry, have been greatly enriched by tests based on the
condition that the cosmology admit a consistent and
observationally acceptable theory for structure formation. The
structure formation theory in turn tests ideas about what the
universe was like before it was well described by the classical
Friedmann-Lema\^{\i}tre model, and may eventually allow us to
enlarge the standard model to include the story of how the world
begins and ends. 

\subsubsection{What is the standard model for structure formation?}

Generally accepted elements are the gravitational growth of
small primeval departures from homogeneity, that may be described
as a stationary isotropic random process, in a universe with
present mass that is dominated by CDM and maybe a term that acts
like a cosmological constant. 

The most striking piece of evidence for the gravitational
instability picture 
is the agreement between the primeval density fluctuations needed
to produce the CBR anisotropy and the present distribution and
motion of the galaxies. Precision measurements in progress should  
allow us to fix many of the details of this gravitational
instability picture, but within present constraints we cannot
say that the primeval density fluctuations are Gaussian, or 
adiabatic, because we have a viable alternative, the
non-Gaussian isocurvature model mentioned in \S~4.

The main piece of evidence for the CDM is the mismatch between
the baryon mass density in the standard model for the origin of
the light elements and the mass density indicated by dynamical
analyses of relative motions of the galaxies. Our reliance on
hypothetical mass is embarrassing; a laboratory demonstration of
its existence would be an exceedingly valuable advance.

\subsubsection{Should we expect surprises from the next generation 
of surveys?}

It is a sign of the growing maturity of our field that we can
pose questions that are motivated by specific theoretical issues
and can be addressed by feasible observations. But I think our
subject still is immature enough that we should be quite prepared
for surprises. My favorite example is 
Shaver's (1991) demonstration that the radio galaxies within
$50h^{-1}$~Mpc distance are close to the plane of the Local
Supercluster, even though the plane is not
apparent in the general distribution of galaxies at
this depth. If the clusters and radio sources were produced by a
pancake collapse why do we not see it in the general galaxy
distribution? Maybe a better picture is that in the early 
universe a nearly straight cosmic string passed by, 
piling mass in its wake into a sheet that
fragmented into the seeds of engines of active galaxies.

I think the most surprising outcome of the new
surveys would be that there are no major corrections to what we
think we know.

\section{Acknowledgments}

This work was supported in part by the USA National Science
Foundation.

\end{document}